\documentstyle[12pt]{article}
\newcommand{\Tr}{\mbox{\rm Tr}}
\begin{document}
\thispagestyle{empty}
\begin{flushright}
{BGUPHYS-TP 95/091}
\end{flushright}
\begin{flushright}
{FIAN/TD-17/95}
\end{flushright}
\begin{flushright}
{August 1995}
\end{flushright}

\begin{center}
{\bf QCD HEAT KERNEL IN COVARIANT GAUGE}
\end{center}

\medskip

\begin{center}
{\bf E.I.~Guendelman${}^a$, A.V.~Leonidov${}^b$,
     V.A.~Nechitailo${}^{a,b}$,  D.A.~Owen${}^a$}
\end{center}

\begin{center}
${}^a${\it Physics Department, Ben Gurion University,
           84105 Beer Sheva, Israel}
\end{center}

\begin{center}
${}^b${\it P.N.~Lebedev Physics Institute,
           117924 Moscow, Russia}
\end{center}

\medskip

\begin{center}
{\bf Abstract}
\end{center}

We report the calculation of the fourth coefficient in an expansion of the
heat kernel of a non-minimal, non-abelian kinetic operator in an
arbitrary background gauge in arbitrary space-time dimension.
The fourth coefficient
is shown to bring a nontrivial gauge dependence due to the contribution
of the lowest order off-shell gauge invariant structure.

\newpage

\begin{center}
{\bf Introduction}
\end{center}

In this letter we continue a program of studying the quasiclassical expansion
of the heat kernel of the gluon kinetic operator considered in the arbitrary
space-time dimension and arbitrary covariant gauge started in \cite{GLNO}.
The importance of studying the heat kernel and its quasiclassical expansion
is due to the important role it plays in quantum field theory and
mathematical physics. In quantum field theory a heat kernel is a unique
device for computing the Green functions and the one-loop quantum corrections
to the classical action of the theory (effective action) in the case of an
inhomogeneous background field. This, in turn, helps to perform an efficient
calculation of charge renormalization and anomalies \cite{RSW}. The method
also provides a possibility of performing an infinite resummation of the
quasiclassical expansion and the corresponding contributions to the effective
action \cite{LZ}. In mathematical physics the heat kernel is a central
object in the spectral geometry of manifolds. Its quasiclassical
expansion generates the invariants of the manifold with respect to the
symmetry transformations of fields and gauge connections defined on it
(see e.g. \cite{Hart}).

Physically the heat kernel is a way of describing the propagation of the
eigenmodes of the system under consideration. As is well known,
the particular
feature needed to determine a non-abelian theory is the necessity
of getting rid
of the unphysical degrees of freedom by fixing a gauge. This leads to a
gauge-dependent description of the propagation of the physical
degrees of freedom,
and, correspondingly, to a gauge dependent heat kernel $K(s)$
\begin{equation}
K(s)=exp(-{\cal W}s),
\label{Kexp}
\end{equation}
where ${\cal{W}}$ is an inverse propagator of a non-abelian gauge boson
(which is actually a covariant Laplacian on the manifold characterized
by the external gauge connections, see below) and $s$ is the proper time.
In the following we shall consider the case of massless gauge bosons
(QCD gluons). The nontrivial application of the heat kernel arises when
one considers a propagation of the quantum gluons on the manifold
characterized by external gauge connections (external non-abelian fields).
The trace of a
heat kernel depends then on the invariants constructed from these external
fields and the question to be studied is the interplay between the gauge
dependence of a heat kernel and various structures (invariants) appearing
in its expansion. As it follows from the results presented below, the
resulting structure is nontrivial.

\begin{center}
{\bf{Calculational Method}}
\end{center}

 We begin with introducing the basic notations. To do this, let us
consider a second order elliptic operator ${\cal W}$ on the
$2\omega$-dimensional manifold ${\cal M}$. By definition the heat kernel
operator corresponding to ${\cal W}$ is obtained by its
exponentiation as given in Eq.~(\ref{Kexp}).
We shall be interested in the expansion of the trace
of the matrix elements of the heat kernel taken at one space-time point.
For the second order operator this expansion has the form (see, e.g.,
\cite{RSW}):

\begin{equation}
\Tr <x|K(s)|x> = \sum_{k=0}^{\infty}
                 b_{-\omega+k}({\cal W},\omega|x)s^{-\omega+k}
\label{TrK}
\end{equation}

\noindent
where the coefficients $ b_{-\omega+k}$ are the so-called Seeley
coefficients. These coefficients are the invariants of the manifold
$\cal{M}$ and the trace of the heat kernel can be considered as a
generating function for these invariants.

For the specific case of a non-abelian gauge theory the kinetic
operator for gluons propagating in the external non-abelian field and
considered in the arbitrary covariant background gauge has the form
\begin{equation}
{\cal{W}}^{ab}_{\mu \nu}=
-D^2(A)^{ab} \delta_{\mu \nu} - 2 f^{acb} G^{c}_{\mu \nu}
-({1 \over \alpha} - 1) D^{ac}_{\mu} D^{cb}_{\nu},
\end{equation}
where $\alpha$ is a gauge fixing parameter, $f^{abc}$ are the structure
constants of the corresponding Lie algebra, $D_{\mu}^a$ is a covariant
derivative containing the external field potential $A_{\mu}$ and
$G_{\mu \nu}$ is the corresponding field strength. The Seeley coefficients
are thus, generally speaking, also the functions of a gauge parameter
\begin{equation}
\Tr <x|e^{-{\cal{W}}^{ab}_{\mu \nu} s}|x> = \sum_{k=0}^{\infty}
                 b_{-\omega+k}(G_{\mu \nu},\omega,\alpha|x)s^{-\omega+k}
\end{equation}
and are invariant with respect to the gauge transformations of the
external field potentials $A_{\mu}^a$.
To calculate the functional trace in (\ref{TrK}) we shall use
the basis of plane waves (~see, e.g., \cite{GolA},\cite{DPYu}~):
$$
\Tr\ <x|e^{-{\cal W}s}|x> = \Tr  \int \frac{d^{2\omega}p}
 {(2\pi)^{2\omega}}\ e^{-ipx}
<x|e^{-{\cal W}s}|x>e^{ipx},
$$

\noindent
where the trace is performed over the Lorentz and color
indices. The exp$[ipx]$ should be pushed through the operator to the
left then cancelled by exp$[-ipx]$. This has the effect that all
differentiation operators in ${\cal W}$ becoming shifted :
$\partial_{\mu} \rightarrow \partial_{\mu} + ip_{\mu}$.
Thus we have

\begin{equation}
  \Tr\ e^{-{\cal W}s} = \Tr \int \frac{d^{2\omega}p}{(2\pi)^{2\omega}}\
e^{-s{\cal W}( \partial_{\mu} \rightarrow \partial_{\mu} + ip_{\mu} )}1,
\label{TrW}
\end{equation}

\noindent
where the operator in the right hand side of (\ref{TrW}) acts on 1.
 For calculational purposes it is convenient to separate  this operator
into the parts having zero, first and second order in the external field
correspondingly:
$$
{\cal W}( \partial_{\mu} \rightarrow \partial_{\mu} + ip_{\mu})
= {\cal W}_{0}-i{\cal W}_{1}-{\cal W}_{2},
$$
where
\begin{eqnarray}
{\cal{W}}^{\nu \nu}_0 & = & p^2 (P^{\mu \nu}_{\bot} +
 {1 \over \alpha} P^{\mu \nu}_{\|}), \nonumber \\
P^{\mu \nu}_{\bot} & = & \delta^{\mu \nu} - {p^{\mu} p^{\nu} \over p^2},\
P^{\mu \nu} = {p^{\mu} p^{\nu} \over p^2}, \nonumber \\
{\cal{W}}_{1}^{\mu \nu} & = & 2 p_{\alpha} D_{\alpha} \delta^{\mu \nu}+
\beta(p^{\mu} D^{\nu}+p^{\nu} D^{\mu}), \nonumber \\
{\cal{W}}_{2}^{\mu \nu} & = & D^2 \delta^{\mu \nu} +2 G^{\mu \nu}
+ \beta D^{\mu} D^{\nu},\ \beta \equiv {1 \over \alpha} - 1
\end{eqnarray}
In the above expression we have suppressed the color indices, which
could be trivially restored, for example
$G_{\mu \nu} \rightarrow G_{\mu \nu}^{ab} = f^{acb} G^c_{\mu \nu}$.

        To obtain the expansion of this operator in $s$ we
use ordinary perturbation theory:
\begin{eqnarray}
\Tr e^{(-{\cal W}_{0}+i{\cal W}_{1}+{\cal W}_{2})s}
 = \Tr K_{0}(s) + \Tr(K_{0}(s)(i{\cal W}_{1}+{\cal W}_{2})) +
\nonumber \\
\int^{s}_{0}ds_{1}(s-s_{1}) \Tr( K_{0}(s-s_{1})
(i{\cal W}_{1}+{\cal W}_{2})K_{0}(s_{1})(i{\cal W}_{1}+{\cal W}s_{2}) )
 + \ldots
\end{eqnarray}
where $K_{0}(s)$ is the free propagator
\begin{equation}
K_{0}(s)=e^{-{\cal W}_{0}^{\mu \nu}s} =
e^{-s p^2} (P^{\mu \nu}_{\bot} + e^{-s \beta p^2} P_{\|}^{\mu \nu})
\end{equation}

The expressions for the Seeley coefficients are obtained by collecting
the terms of a given order in covariant derivatives.

In the next section we shall apply the above described method to calculate
the fourth Seeley coefficient for the gluon kinetic operator
in the background $\alpha$ - gauge.
\newpage

\begin{center}
{\bf Fourth Seely Coefficient in $\alpha$-gauge }
\end{center}

We start with recalling the first three Seely coefficients \cite{GLNO}. They
have the following form :
\begin{eqnarray}
b_{-\omega} & = & \frac{N^2_c-1}{2^{2\omega}\pi^{\omega}}
              (2\omega - [1-\alpha^{\omega}]),
     \label{b0}
     \\
b_{-\omega+1} & = & \frac{N_c}{2^{2\omega}\pi^{\omega}}
              (2\omega - [1-\alpha^{\omega-1}])
 \frac{\Gamma(\omega)-\frac{1}{\omega}\Gamma(\omega)}{\Gamma(\omega)}
 A^a_\mu A^a_\mu \equiv 0 ,
       \\
b_{-\omega+2} & = & \frac{N_c}{2^{2\omega}\pi^{\omega}}
               (2\omega - [1-\alpha^{\omega-2}])
               G^a_{\mu\nu} G^a_{\mu\nu} .
       \label{b2}
\end{eqnarray}

The fourth Seeley coefficient corresponds to collecting the contributions
of the sixth order in the covariant derivatives. In this order there
exist two gauge invariant structures. These are
$$
 G_3 \equiv f^{abc}G^a_{\mu\nu} G^b_{\nu\rho} G^c_{\rho\mu}, \ \ \
I_3 \equiv (D^{ab}_\mu G^b_{\mu\nu})( D^{ac}_\rho G^c_{\rho\nu}).
$$
The second invariant $I_3$ is nonzero only for the fields that do not
obey the classical equations of motion, i.e. off-shell. The expression
for the fourth Seeley coefficient $b_{-\omega+3}$ reads
\begin{equation}
b_{-\omega+3} = - \frac{N_c}{2^{2\omega}\pi^{\omega}}
 \left[\frac{1}{180}(2\omega - [1-\alpha^{\omega-3}])(G_3 - I_3)
 + \frac{2}{3}I_3 + \xi(\omega,\alpha) I_3 \right]
\end{equation}
where
\begin{eqnarray}
\xi(\omega, \alpha) = \frac{1}{12\omega(\omega-1)(\omega-2)} \left\{
 10\omega^2 -25\omega+6 - 6\frac{\alpha}{1-\alpha}(2\omega-1)
 \right. \nonumber\\
 \left.    + \frac{\alpha^{\omega-2}}{1-\alpha}
 [2\omega(\omega-1)+\omega(11-2\omega)\alpha+3(\omega-2)\alpha^2]
 \right\}
                  \label{xi}
\end{eqnarray}
This expression is the main result of our paper.
We see that it is indeed gauge invariant. The most
interesting property of the above expression is its dependence
on the gauge fixing for quantum gluons, i.e. its $\alpha$ dependence.
First of all, one immediately sees, that in the limit
$\alpha \rightarrow 1$ (Feynman gauge)
%
% vstavka
%
the function $\xi(\omega,1)$ equals  zero identically for any $\omega$
and for $\omega = 2$ the remaining part of $b_1$ coincides with
that obtained in \cite{DPYu} in this particular case.
Let us note that function $\xi(\omega,\alpha)$ also does not have poles
at any $\omega$ and the corresponding limits are
\begin{eqnarray}
\xi(0,\alpha) & = & - \frac{1}{24}\{ \frac{2}{\alpha^2}-\frac{9}{\alpha} + 13
                    + 6 \frac{\ln(\alpha)}{(1-\alpha)}\},
                \label{om0}
                \\
\xi(1,\alpha) & = & - \frac{1}{12}\{ 2( \frac{1}{\alpha}+2 )
                    + 3 \frac{\ln(\alpha)}{(1-\alpha)}(3-\alpha) \},
                  \label{om1}
                  \\
\xi(2,\alpha) & = &   \frac{1}{12}\{ 3(7-\alpha) +
                       2 \frac{\ln(\alpha)}{(1-\alpha)}(2+7\alpha) \}.
                  \label{om2}
\end{eqnarray}
 From a physical point of view the expressions (\ref{om0}) and (\ref{om1})
do not make any sense but they show that "explicit" poles in (\ref{xi})
are not the actual poles. In fact the function $\xi(\omega, \alpha)$
reflects a non-trivial off-shell contribution to $b_{-\omega+3}$.
Its dependence on $\alpha$ and $\omega$ is complicated and
at this moment we do not have any further comments on its structure.

 The second important property of the coefficient $b_{-\omega+3}$
is that the on-shell contribution, which is proportional to $G_3$
is proportional to the factor $2\omega-[1-\alpha^{\omega-3}]$.
A comparison with the expressions for other coefficients
(~Eqs.~(\ref{b0})-(\ref{b2})~)
shows, that
 the on-shell contribution to the heat kernel in order  $n$ contains a
universal factor $2\omega - [1-\alpha^{\omega-n}]$.
A very plausible conjecture
is that this is valid for all terms in the heat kernel expansion.
As was shown in \cite{GLNO}, this corresponds to the appearance of the
contributions proportional to the $\ln \alpha$ in the $\zeta$ - regularized
gluon contribution of the effective action.

\begin{center}
{\bf Conclusion}
\end{center}

Following the line of research described in our previous paper \cite{GLNO}
we have calculated the fourth coefficient in the quasiclassical expansion
for the trace of a heat kernel operator for the kinetic operator for
a non-abelian gauge boson in an arbitrary background gauge and arbitrary
space-time dimension. The new coefficient is of the sixth order in the
covariant derivatives with respect to the background field gauge potentials
and is the first one containing two different invariants with respect to
gauge transformations of the background field. One of the invariants is an
on-shell one
(~$G_3 \equiv f^{abc}G^a_{\mu\nu} G^b_{\nu\rho} G^c_{\rho\mu}$~),
and another one
(~$I_3 \equiv (D^{ab}_\mu G^b_{\mu\nu})( D^{ac}_\rho G^c_{\rho\nu})$~) is
an off-shell one, i.e. vanishes for the external fields obeying classical
equations of motion. The structure of the coefficient corresponding to the
on-shell contribution is proved to follow the pattern seen in the previously
calculated coefficients (see \cite{GLNO}). The off-shell contribution is
found to be a complicated function of the quantum gauge fixing parameter
and the space-time dimension.

The results obtained in this paper are of considerable importance  for
the analysis of the general structure of the contributions to the effective
action. In particular, it is important to understand the structure of
the heat kernel
expansion and the corresponding contributions to the effective action at
finite temperature (i.e. on the cylindrical space-time manifold).

\begin{center}
{\bf Acknowledgements}
\end{center}
The work of A.~L. was supported by the Russian Fund for Fundamental Research,
Grant 93-02-3815. V.~N. is grateful to A.P.~Nechitailo for help in
carrying out computer calculations. The work of V.~N. was supported
in part by Soros Foundation, Grant M5V000.

\end{document}